\documentstyle{article}

\begin{document}
\title{Entanglement of unstable atoms: modifications of the emission properties}
\author{ P. Sancho (a), L. Plaja (b)  \\ (a) Delegaci\'on de AEMET en
Castilla y Le\'on. \\ Ori\'on 1, 47014, Valladolid, Spain \\
(b) Area de Optica. Departamento de F\'{\i}sica Aplicada.
Universidad
\\ de Salamanca. Pl. de la Merced s/n, 37008, Salamanca, Spain }
\date{}
\maketitle
\begin{abstract}

We analyse the influence of entanglement on the emission properties
of atoms. To this end, first, we propose a scheme for the
preparation of a pair of entangled Helium atoms, one in the ortho
and the other in the para spin configuration. We discuss a realistic
scenario for this process, based in the double ionization of He by
intense laser fields. These states are used to analyse
disentanglement  and the role of entanglement in the spontaneous
emission from the pair. In particular, we show that the decaying
rate of an entangled atom is different from that in a product state,
modifying the temporal emission distribution and lifetime of the
atoms.
\end{abstract}
\vspace{7mm}


\section{Introduction}
Entanglement has become a central issue in quantum theory. Many
aspects of the subject have been already deeply analysed as, for
instance,  questions about its experimental generation and
verification \cite{NC, van, aud}, the existence of good measures for
mixtures \cite{ple}, its application in quantum information
processing and foundational tests \cite{NC}, and many others.

More recently, several authors have studied the behavior of
entanglement during the process of spontaneous emission. Di\'osi
\cite{diosi} and  Dood {\em et al} \cite{dodd} showed that
entanglement is destroyed by the same decoherence mechanisms that
degrade quantum interference. In a fundamental paper \cite{Ebe} the
disentanglement dynamics originated by the spontaneous emission
process was analyzed (see also \cite{Yu,Jag} and references
therein). In that approach it was demonstrated that, in the general
case, entanglement decays at least as fast as the sum of decaying
rates of the individual states. Also the entanglement existent
between an atom and a spontaneously emitted photon has been
evaluated \cite{Fed}. Using the Schmidt decomposition \cite{jpb},
that entanglement can be estimated in a quantitative way. The same
approach can be extended to other decaying systems, even when there
is interaction between the products of the process \cite{fp}. In a
related context it has been proposed in Ref. \cite{chi} to use
squeezed fields to preserve entanglement against spontaneous
emissions.

In this paper we want to consider a less known aspect of the problem, the influence of entanglement on the
emission properties of unstable atoms. This aspect is complementary
to those studied in Refs. \cite{Ebe, Fed}. We analyze the problem by considering a system particularly well suited for
this purpose, an entangled state of excited ortho- and para-Helium
atoms. Recently, it has been suggested the possibility of preparing
metastable superpositions of ortho- and para-Helium states
\cite{SP}. We shall begin this paper addressing the problem of state
preparation, and  we shall discuss the interaction with intense
laser filed as a possible scenario in which this entanglement
appears naturally.

Then we will proceed to study radiative properties by calculating
the decay rates of the entangled pair in terms of those of
unentangled atoms. Three main conclusions will be obtained. First,
we shall derive the decaying rate of the entangled state using a
simple general method that does not resort to the master equation of
the system (as done in Ref. \cite{Ebe}). Later, and as the most
important result of the paper, we show that the decaying rates of
entangled and free unstable atoms are different. The amplitude of
this effect is measured by expectation values of the non-trivial
part of the evolution operator. Finally, being the decaying rates
different, we can expect the temporal emission distributions and
lifetimes also to be different. We confirm this behavior by
calculating the emission distributions in entangled and product
states. The results presented here aim to provide a deeper
understanding of some fundamental properties of entangled unstable
states, that are in principle accessible to experimental scrutiny.

We must remark the resemblances and connections of our work with
other related problems that have already been addressed in the
literature. In the field of quantum information, the authors in
\cite{Pal,Rei} have analyzed the decay of the density matrix
coherences of one or several qubits, used as quantum registers, when
they are coupled with a bath. The density matrix coherences measure
the degree of superposition between the two states of the qubit.
When the qubit interacts with the bath the state of the complete
system, qubit plus bath, becomes entangled. The decaying rates of
the coherences show a strong dependence on the characteristics of
the entangled state and the bath. The connection of this behaviour
with our main result is clear. At variance with these papers, we
shall not consider the decaying of a superposition state but that of an
excited one. On the other hand, the resemblances of our approach with
the work in \cite{Pur,bra} (and references therein), where the
authors have considered the modifications of the emission properties
associated with the presence of boundaries, will be addressed in the
Discussion section.

\section{Preparation of the state}

In Ref. \cite{SP} it was discussed how to prepare a superposition of
ortho- and para-Helium states. That proposal can easily be extended
to the generation of entangled states. Suppose we have two Helium
ions $He^+$ with their electrons in the $|\uparrow \rangle$ state (in a
predetermined axis, for instance the $z$-one): $|He^+(\uparrow )\rangle_L
|He^+(\uparrow )\rangle_R$. We also assume that both ions are placed at
distant regions denoted by $L$ and $R$. Next, we prepare a pair of
electrons in the entangled state
\begin{equation}
|e_L e_R\rangle=\frac{1}{\sqrt{2}} |e(\uparrow )\rangle_L |e(\downarrow )\rangle_R +
\frac{1}{\sqrt{2}} |e(\downarrow )\rangle_L |e(\uparrow )\rangle_R
\label{eq:uno}
\end{equation}

When there is capture, the interaction between electrons and ions is given by
\begin{equation}
|He^+(\uparrow )\rangle|e(\uparrow )\rangle  \rightarrow |He_{or}\rangle
\end{equation}
and
\begin{equation}
|He^+(\uparrow )\rangle|e(\downarrow )\rangle  \rightarrow |He_{pa}\rangle
\end{equation}
Therefore, because of the linearity of quantum theory we have
\begin{eqnarray}
|He^+(\uparrow )\rangle_L |He^+(\uparrow )\rangle_R |e_L e_R\rangle \rightarrow
\nonumber \\
 \frac{1}{\sqrt{2}} |He_{or}\rangle_L |He_{pa}\rangle_R +
\frac{1}{\sqrt{2}} |He_{pa}\rangle_L |He_{or}\rangle_R \label{eq:cua}
\end{eqnarray}
This is a maximally entangled state of otho- and para-Helium.

In order to actually implement this scheme one must be cautious with
the energies of the incident electrons. We want to prepare
metastable states. The simplest choice is the $1s2s$ states of both
configurations, ortho- and para-Helium: we must prepare state
(\ref{eq:cua}) with $He_{pa}(1s2s)$ and $He_{or}(1s2s)$. The
difference of energy between the lower state of the ortho- ($1s2s$)
and  the first excited state of the para-Helium ($1s2s$) is small
and it seems possible to fulfill that condition.

From now on we shall only refer to these states, using the notation
\begin{eqnarray}
|\Psi \rangle & = & \frac{1}{\sqrt{2}} |He_{or}(1s2s)\rangle_L |He_{pa}(1s2s)\rangle_R + \label{eq:cin}
\frac{1}{\sqrt{2}} |He_{pa}(1s2s)\rangle_L |He_{or}(1s2s)\rangle_R  \nonumber \\
& = & \frac{1}{\sqrt{2}} |or\rangle_L |pa\rangle_R + \frac{1}{\sqrt{2}} |pa\rangle_L |or\rangle_R
\end{eqnarray}

In the next section we discuss two ways to actually implement the above scheme.

\section{Implementation of the scheme}

Interestingly, a situation similar to the above procedure takes
place during the interaction of strong laser radiation with a cloud
of He atoms. It is known that, at intensities near saturation, the
most probable path for the double ionization of He is the so-called
sequential process \cite{walker}. In this case, the two electrons
are ionized at each of the field maxima of a laser cycle. Since the
field interaction is not relativistic, the ionized electron pair
retains the same spin configuration as the He ground state. In
addition, as the field reverses its sign during the laser cycle, the
electrons are emitted to opposite sides of the parent atom,
configuring a spatially separated entangled pair. On the other hand,
the single ionization of He occurs typically with a higher rate than
the double. Therefore, the ionized pair has higher probability to
encounter single ionized atoms than double ionized. As a
consequence, there is a better chance to recombine in the form of
entangled pairs of neutral He than of He$^+$. Half of these neutral
recombinations will produce He entangled pairs of the form given in
Eq. (\ref{eq:cua}). The final yield  depends on the particular cross
section of the recombination process. However note that, as we are
considering low-energy metastable states in our discussion, their
fraction will increase as other (unwanted) configurations will decay
via spontaneous processes.

\section{Lifetimes of entangled states}

Our starting point is Eq. (\ref{eq:cin}). The states $1s2s$ of
ortho- and para-Helium atoms are metastable. Therefore, after some time
they will decay emitting photons. As the energies of the photons
emitted by ortho and para atoms are (slightly) different, we can in principle
know if the emission event took place for a para- or ortho-atom and,
consequently, if the companion atom rests in an ortho- or para-state.
Moreover, since the two particles are spatially well separated, the
events to emit in $L$ or to emit in $R$ are distinguishable. We can
know where and by what type of state was the photon emitted. The two
alternatives in Eq. (\ref{eq:cin}) become distinguishable and there
is no longer a superposition of two-particle states. The pure
two-particle state (\ref{eq:cin}) disentangles to a mixture of
states
\begin{equation}
|g\rangle_i |or\rangle_j \; ,
|g\rangle_i |pa\rangle_j \;
\end{equation}
with $i,j=L,R$, $i \neq j$ and $|g \rangle$ denoting the ground state, i. e.,
the para-Helium $1s1s$ state at which the two $1s2s$ states decay.

We can ask now how fast this process is or, equivalently, by the
lifetime of the entangled state (\ref{eq:cin}). This property can be
easily evaluated using the standard techniques for the calculation
of lifetimes of unstable atoms. The lifetime of (\ref{eq:cin}),
denoted as $\tilde{\tau }$, is given by $\tilde{\tau
}=1/\tilde{\Gamma }$ with $\tilde{\Gamma }$ the decaying rate of
$|\Psi \rangle$. From now on in order to avoid any confusion between
the decaying rates and lifetimes of entangled and non-entangled

states, those corresponding to the first case will be denoted by a
tilde.

The decaying can occur at two different locations,
\begin{equation}
|\Psi \rangle \rightarrow |g\rangle_L |h\rangle_R
\end{equation}
at the left side, and
\begin{equation}
|\Psi \rangle \rightarrow |h\rangle_L |g\rangle_R
\end{equation}
at the right side. $|h\rangle$ can have two different values at each side, $h=or,pa$.

Since we add probabilities and not amplitudes (because, as signaled
before, the alternatives of emission at $L$ or at $R$ are
distinguishable) we have that the total decaying rate is
\begin{equation}
\tilde{\Gamma }= \tilde{\Gamma }_L + \tilde{\Gamma } _R
\end{equation}
where $\tilde{\Gamma }_L$ and $\tilde{\Gamma } _R$ denote the decaying
rates of the two channels associated with the two above equations.

Let us evaluate $\tilde{\Gamma }_i$, $i=L,R$. Taking into account
again that the emissions by atoms ortho and para are distinguishable
and, consequently, we must add probabilities instead of amplitudes
we have that the probability of the decaying process per unit time
in the interval $(t_0,t)$ is
\begin{equation}
\tilde{\Gamma }_i (t,t_0)=\frac{1}{t-t_0} \sum _{|h\rangle_j} \; \; | _j\langle h| _i\langle g|\hat{U}(t,t_0)|\Psi \rangle|^2
\label{eq:DiE}
\end{equation}
with $h=or,pa$ and $i,j=L,R ; i \neq j$.

When we take the limit of $t-t_0$ very small we obtain the
instantaneous value of the decaying rate. This justifies the use of
the same notation, $\tilde{\Gamma }$, for the instantaneous decaying rates
and for this probability per unit time.  $\hat{U}(t,t_0)$ is the
evolution operator between times $t$ and $t_0$. It is given by $\exp
(-i\hat{\cal H}(t-t_0))$ with $\hat {\cal H}$ the Hamiltonian
operator of the system. We can decompose the Hamiltonian in the form
$\hat{\cal H}=\hat{\cal H}_L+\hat{\cal H}_R$ because there is not
interaction between the well separated parts $L$ and $R$ (note that
$\hat{\cal H}_L = \hat{\cal H}_R$ except by the fact that they act
in different spatial regions). Therefore, since $[\hat{\cal H}_L,
\hat{\cal H}_R]=0$ the evolution operator factorizes as
$\hat{U}=\hat{U}_L \hat{U}_R $.

Now, we are in position
to evaluate the disentangling rate per unit time in the interval
$(t_0,t)$:
\begin{eqnarray}
\tilde{\Gamma }(t,t_0)= \tilde{\Gamma }_{L}(t,t_0) + \tilde{\Gamma }_{R}(t,t_0) = \nonumber \\
\tilde{\Gamma }_{or}^L(t,t_0) + \tilde{\Gamma }_{pa}^L(t,t_0)+ \tilde{\Gamma }_{or}^R(t,t_0) + \tilde{\Gamma }_{pa}^R(t,t_0) =   \nonumber \\
\tilde{\Gamma }_{or}(t,t_0) + \tilde{\Gamma }_{pa}(t,t_0)
\end{eqnarray}
Finally, for the lifetime we have
\begin{equation}
\tilde{\tau } \approx \frac{1}{\tilde{\Gamma _{or}}+\tilde{\Gamma _{pa}}}
\label{eq:caca}
\end{equation}

Equation (\ref{eq:caca}) shows that the lifetime of the entangled state is always shorter than those of
the component states. In our case, it is slightly shorter than that
of the $1s2s$ state of the para-Helium. This result agrees with that obtained in \cite{Ebe}, where it was
shown that entanglement decays at least as fast as the sum of the
separate rates.

The result can easily be understood in terms of the probabilities
associated with the two decay channels. As the two channels
represent distinguishable alternatives the probability of
disentanglement is the sum of the two decay probabilities and,
consequently, larger than any of them.

\section{Decaying rates of atoms in entangled states}

In this section we analyze the relation between entangled and
unentangled decaying rates. Let us see, for instance, the case
$h=or$ in Eq. (\ref{eq:DiE}), which corresponds to the evolution
$|\Psi \rangle \rightarrow |g\rangle_i |or \rangle_j$. The
probability amplitude of this process in the interval $(t_0,t)$ is
\begin{eqnarray}
_j\langle or |_i\langle g|\hat{U}(t,t_0)|\Psi \rangle =\frac{1}{\sqrt{2}} \, _j\langle or |\hat{U}_j(t,t_0)|pa\rangle_j
\, _i\langle g|\hat{U}_i(t,t_0)|or \rangle_i  \nonumber \\
+ \frac{1}{\sqrt{2
}}\,  _j\langle or|\hat{U}_j(t,t_0)|or\rangle_j \,
_i\langle g|\hat{U}_i(t,t_0)|pa\rangle_i = \nonumber \\
\frac{1}{\sqrt{2}} {\cal M}_{op}^j {\cal M}_{go}^i +
\frac{1}{\sqrt{2}} {\cal M}_{oo}^j {\cal M}_{gp}^i
\end{eqnarray}
where the notation for the matrix elements ${\cal M}$ is obvious.

It must be noted the presence of the term ${\cal M}_{op}^j {\cal
M}_{go}^i$. Physically, it represents the emission by an atom ortho
at $i$ (leaving the atomic state $|g\rangle_i$) and the transition $|pa\rangle
\rightarrow |or\rangle$ at $j$ (giving rise to the presence of an unstable
atom ortho at $j$). This term contributes to the total decaying rate
of the channel $|\Psi \rangle \rightarrow |g\rangle_i |or \rangle_j$,
but not to the
emission rate of photons emitted from the para configuration.

The probability of this transition is extremely small and,
consequently, we shall neglect this contribution here. With this
approximation the only relevant contribution to the disentangling
channel $|\Psi \rangle \rightarrow |g\rangle_i |or \rangle_j$ is
${\cal M}_{oo}^j {\cal M}_{gp}^i$. The physical meaning of these
matrix elements is clear. $|{\cal M}_{gp}^i|^2/(t-t_0) $ gives the
probability of decaying of a free unstable para-type atom (located
at $i$) per unit time in the interval $(t_0,t)$, $\Gamma
_{pa}^i(t,t_0)$. As discussed before, in the limit of $t-t_0$ very
small it gives the instantaneous decaying rate of free unstable
para-type atoms at side $i$, $\Gamma _{pa}^i(t_0)$. It also equals
the emission rate of photons emitted from the para configuration at
$i$.

The other matrix element, ${\cal M}_{oo}^j$, can be
expressed in the well-known way
\begin{equation}
{\cal M}_{oo}^j = \, _j \langle or|1 + \hat{W}_j (t,t_0) |or\rangle_j=1+W_j^{or}(t,t_0)
\end{equation}
The operator $\hat{W}_j$ contains all the terms in the expansion of
the evolution operator depending on the Hamiltonian, i. e., it represents
the non-trivial terms. On the other hand, $W_j^{or}(t,t_0)= \, _j \langle or |
\hat{W}_j(t,t_0) |or\rangle_j $ is the expectation value of the operator
$\hat{W}_j(t,t_0)$ in the ortho state.

Thus, the squared modulus of ${\cal M}_{oo}^j {\cal M}_{gp}^i
/\sqrt{2}$ divided by $t-t_0$ gives the probability of
disentanglement by emission of a photon of type para at side $i$.
Moreover, it equals the emission rate by an unstable para-atom in an
entangled state at side $i$. Taking into account this equality, we
use the same notation for the disentanglement and decaying for
unstable entangled atom rates:
\begin{equation}
\tilde{ \Gamma } _{pa}^i (t,t_0)=\frac{1}{2} \Gamma _{pa}^i (t,t_0) (1+|W_j^{or}(t,t_0)|^2+ 2Re(W_j^{or}(t,t_0)))
\end{equation}

The total decaying rate per unit time of atoms of type para in an
entangled state is
\begin{equation}
\tilde{ \Gamma } _{pa} (t,t_0)=\tilde{ \Gamma } _{pa}^L (t,t_0) + \tilde{ \Gamma } _{pa}^R (t,t_0)= \Gamma _{pa}(t,t_0) (1+|W^{or}(t,t_0)|^2+ 2Re(W^{or}(t,t_0)))
\end{equation}
In the last term we have dropped all the indexes referring to the
location of the atom ($j=L,R$) because all the variables are equal
on both sides due to the symmetry of the problem.
A similar result can be obtained for the ortho case with obvious
modifications.

We have obtained (in the limit of $t-t_0$ very
small) the decaying rates of atoms in entangled states.
This expression clearly shows that the decaying rate of an atom in
an entangled state differs from that in a product state due to the
presence of $W^h$, $h=or,pa$.

 As we shall see in next section this property has
interesting physical implications. The importance of
these differences is measured in the para case by
\begin{equation}
\frac{\tilde{\Gamma }_{pa}(t-t_0) -\Gamma _{pa}(t-t_0)}{\Gamma _{pa}(t-t_0)}=|W^{or}(t,t_0)|^2+ 2Re(W^{or}(t,t_0))
\end{equation}

The disentangling rate per unit time in the interval
$(t_0,t)$ can be expressed in terms of the unentangled rates as:
\begin{equation}
\tilde{\Gamma }(t,t_0)= \Gamma _{or}(t,t_0) + \Gamma _{pa}(t,t_0) + \Lambda (t,t_0)
\end{equation}
with
\begin{equation}
\Lambda (t, t_0)= |W^{or}(t,t_0)|^2+ |W^{pa}(t,t_0)|^2+
2Re(W^{or}(t,t_0) + W^{pa}(t,t_0))
\end{equation}
In the case that the term $\Lambda $ is very small (using the limit
$t \rightarrow t_0$ and assuming a stationary situation where the
decaying rates are time independent) we have
\begin{equation}
\tilde{\tau } \approx \frac{1}{\Gamma _{or}+\Gamma _{pa}}
\end{equation}

\section{Distributions of emitted photons. Lifetimes}

As discussed in the previous section the decaying rates of entangled
atoms differ from those of free unstable ones. Then one can expect
the temporal distributions of emitted photons and lifetimes of unstable atoms do not be
equal in both cases. We confirm this point by an explicit
calculation of the time-dependent distributions.

If we denote by $n$ and $n_h$ respectively the number of pairs of
entangled particles and non-entangled unstable atoms (the unstable
atoms produced in the disentangling process) of type $h$, taking the
usual approach of an exponential decay, their rules of change are
given by
\begin{equation}
\frac{dn}{dt}=-\tilde{\Gamma }n
\end{equation}
and
\begin{equation}
\frac{dn_h}{dt}=\tilde{\Gamma }_H n -\Gamma _h n_h
\end{equation}
where $h,H=or, pa$ and $H \neq h$. Along this section we use the
instantaneous emission rates and assume, by simplicity, that they
are independent of time. The first equation says that the number of
entangled states only varies because of disentanglement. The rule
for $n_h$ is different. We have a source term $\tilde{\Gamma }_H n $
that gives the number of entangled pairs that generate an unstable
atom of type $h$ (the emitted photon in the disentangling process is
of type $H$). On the other hand, we have the usual decay term,
$\Gamma _h n_h$. The above equations can easily be solved. Taking
the initial conditions $n(0)=n_0$ (with $n_0$ the initial number of
entangled pairs) and $n_h(0)=0$ (initially all the atoms are
entangled) we have
\begin{equation}
n(t)=n_0 e^{-\tilde{\Gamma }t}
\end{equation}
and
\begin{equation}
n_h(t)=n_0 \frac{\tilde{\Gamma }_H}{\tilde{\Gamma } - \Gamma _h}
(e^{-\Gamma _h t} - e^{-\tilde{\Gamma }t})
\end{equation}
To solve the second equation we have taken for the particular
solution $\propto \, e^{-\tilde{\Gamma }t}$ and for the general of the
homogeneous equation $\propto \, e^{-\Gamma _h t}$.

These expressions are to be compared with that obtained for product states, which is derived from
\begin{equation}
\frac{dn_h^p}{dt}=-\Gamma _h n_h^p
\end{equation}
with solution
\begin{equation}
n_h^p(t)=n_0 e^{-\Gamma _h t}
\end{equation}
where we have denoted all the variables related to the product state by the superscript $p$.

Clearly, $n_h(t)$ and $n_h^p(t)$ are, in general, different. The
most interesting manifestation of these different behaviours is,
probably, the fact that the lifetimes are not equal. In effect, the
lifetime of a free (or in a product state) unstable atom of type
$h$, $\tau _h$, is defined by the relation
\begin{equation}
n_h^p(\tau _h)=n_0 e^{-1}
\end{equation}
This definition can be extended to the case in which the atoms are
initially entangled. In this case, however, one must be cautious
because the unstable atoms of type $h$ (before of decaying to the
stable state) can be in an entangled or in a product state. The
number of atoms in both types of states must be added. Then the
definition of the lifetime of the atom of type $h$ when initially is
in an entangled state, $\tilde{\tau }_h$, is
\begin{equation}
n(\tilde{\tau }_h)+ n_h(\tilde{\tau }_h)=n_0 e^{-1}
\end{equation}
because the number of unstable atoms of type $h$ in an entangled state at
time $t$ equals the number of entangled pairs at that time, $n(t)$.

Then $\tilde{\tau }_h$ is given by the solution of the equation
\begin{equation}
\left( 1- \frac{\tilde{\Gamma }_H}{\tilde{\Gamma }- \Gamma _h} \right)
e^{-\tilde{\Gamma }\tilde{\tau }_h} +  \frac{\tilde{\Gamma }_H}{\tilde{\Gamma }- \Gamma _h}
e^{-\Gamma _h \tilde{\tau }_h} = e^{-1}
\end{equation}

From an operative point of view it is not easy to determine
experimentally the number of atoms of each type existent at a given
time. It is much easier to measure the number of photons of each
type emitted at that time, and from this data to obtain the number
of unstable atoms. We denote by $N_h(t)$ the number of photons of
type $h$ emitted at time $t$. It is given by
\begin{equation}
\frac{dN_h}{dt}=\tilde{\Gamma }_h n + \Gamma _h n_h
\end{equation}
Using the expressions for $n(t)$ and $n_h(t)$ and the initial
condition $N_h(0)=0$ (in this step we must use $\tilde{\Gamma } =
\tilde{\Gamma }_h + \tilde{\Gamma }_H$) we obtain
\begin{equation}
N_h(t)= n_0 + \frac{n_0}{\tilde{\Gamma }} \left( \frac{\Gamma _h
\tilde{\Gamma }_H}{\tilde{\Gamma }- \Gamma _h} -\tilde{\Gamma }_h \right)
e^{-\tilde{\Gamma }t} + n_0 \frac{\tilde{\Gamma }_H}{\Gamma _h - \tilde{\Gamma }}
e^{-\Gamma _h t}
\end{equation}
This expression differs from the distribution of emitted photons by
an atom of type $h$ when it is in a product state, which is
\begin{equation}
N_h^p (t)=n_0 (1-e^{-\Gamma _h t})
\end{equation}

The number of unstable atoms of type $h$ at any time can be deduced
from the easily measurable distributions of emitted photons when the temporal
distribution is measured on a large number of single repetitions of
the experiment (instead of a large number of pairs prepared at the
same time in the same zone). In effect, then for each pair we can
determine if each photon has been emitted before than the other
(causing the disentangling process) or after it (the emission is
caused by a non-entangled unstable atom). We denote the distribution
of the first type of photons by $N_h^{(1)}(t)$ and that of the
second ones by $N_h^{(2)}(t)$. Then
$n_h(t)=N_H^{(1)}(t)-N_h^{(2)}(t)$. We can also determine $n(t)$
using the relation $n(t)=n_0- N_H^{(1)}(t)-N_h^{(1)}(t)$ In contrast,
when all the pairs are at the same time in the same zone (for
instance, a cloud of initially entangled pairs) one cannot determine
if the photon is of type $(1)$ or $(2)$ and it is not possible to
deduce the distributions $n(t)$ and $n_h(t)$.

It must be signaled as a byproduct that the above emission
distributions could be used to determine without additional
measurements if a state is entangled or not. This type of
characterization of entanglement is usually denoted as a direct
measurement of entanglement \cite{van}. A similar proposal can be
found in Ref. \cite{Rod}, where the authors use the correlations
function of the emitted light for detecting entanglement in a system
comprising coupled excitonic qubits.

\section{Discussion}

We have shown in this paper that it is, in principle, possible to
prepare metastable entangled states of atoms. Although we have
restricted our considerations to the Helium the same scheme can be
extended to metastable states of other atoms. For states for which
the transition between them can be neglected (as for ortho- and
para-Helium), the extension is immediate. When that condition is not
fulfilled the matrix element representing the transition must be
taken into account. In future work we plan to discuss how to
actually implement the scheme proposed here.

The proposed technique allows to analyze the properties of unstable
atoms in entangled states. Three principal results have been
obtained in this paper: (i) We have derived the disentanglement rate
in a simple and intuitive way that does not resort to more involved
techniques as the master equation used in Ref. \cite{Ebe}. (ii) We
have shown that the decaying rates of the atoms are different for
entangled and product states. (iii) As a consequence of (ii), the
temporal distributions of emitted photons and lifetimes of unstable
atoms are modified by the presence of entanglement.

The results (ii) and (iii) provide a complementary view to the
relation between entanglement and spontaneous emission presented in
\cite{Ebe, Yu, Jag, Fed} (and point (i) here). In effect, in these
references (and the evaluation of
$\tilde{\tau }$ here) one studies how the entanglement
is generated \cite{Fed} or destroyed \cite{Ebe} by spontaneous
emission. The results presented here are complementary to the above
ones, showing that entanglement is not only a passive element, but
also an active one that can modify the emission properties.

The modifications of the emission properties and lifetimes due to the
presence of entanglement provide a
second example of modifications of the properties of unstable atoms.
In \cite{bra} it was shown how the spontaneous emission rate is
altered when placed inside a conducting wedge. Physically, this
effect can be explained because the non-trivial boundary conditions
affect the interaction between atoms and the quantum electromagnetic
field. In our example the modifications are associated with the
entanglement of the atom with another atom. The wavefunction, with
which we evaluate all the physical properties of the system, is
different from that of a free system.

Finally, it is worthy to examine our results from the point of view
of the reduced density matrix. It is remarkable the fact
that, according to our analysis, we have properties of the
individual system that are modified by  the presence of
entanglement. It is sometimes vaguely stated that only joint properties of the
two-particle system are sensitive to the presence of entanglement.
The reduced (or single-particle) density matrix of any of the atoms
for state (\ref{eq:cin}) has the form $\rho _h = Tr _H(|\Psi
\rangle\langle \Psi |)=\frac{1}{2} |h\rangle_R\langle h| + \frac{1}{2} |h\rangle_L \langle h |$, which is
equivalent to a mixture of states $|h\rangle_R$ and $|h\rangle_L$. The expected
value of any observable $\hat{A}$ associated with the system is
$\langle\hat{A}\rangle=Tr(\rho _h \hat{A})$. Thus, the expectation value of any
observable for the entangled state and the mixture are equal and,
consequently, one cannot use the values of $\langle\hat{A}\rangle$ to
distinguish between both types of states. Nevertheless, the decaying
rates, lifetimes and emission distributions are not expectation
values of observables and are not constrained by the above
restriction.

\end{document}